\documentclass[11pt]{iopart}
\usepackage{citesort}
\usepackage{latexsym}

\newcommand{\curl}{\mathrm{curl}}%

\newcounter{mnotecount}[section]

\renewcommand{\themnotecount}{\thesection.\arabic{mnotecount}}

\newcommand{\mnote}[1]%{}%
{\protect{\stepcounter{mnotecount}}$^{\mbox{\footnotesize
$%\!\!\!\!\!\!\,
\bullet$\themnotecount}}$ \marginpar{%\color{red}%
\raggedright\tiny\em
$\!\!\!\!\!\!\,\bullet$\themnotecount: #1} }

\def\be{\begin{equation}}
\def\ee{\end{equation}}
\def\bea{\begin{eqnarray}}
\def\eea{\end{eqnarray}}
\def\bean{\begin{eqnarray*}}
\def\eean{\end{eqnarray*}}

\newfont{\BlackBoardBold}{msbm10}
\def\s{\sigma}

\begin{document}

\title[Lanczos potentials and gravitational entropy for perturbed FLRW
space-times]{Lanczos potentials and a definition of gravitational
entropy for perturbed FLRW space-times}

\author{Filipe C. Mena$^\flat$ and Paul Tod$^\star$}
\address{$^\flat$Departamento de Matem\'atica,
Universidade do Minho, 4710 Braga, Portugal}
\address{$^\star$Mathematical Institute, University of Oxford, St. Giles 24-29,
Oxford OX1 3LB, UK}

%\date{}
\eads{\mailto{fmena@math.uminho.pt}, \mailto{tod@maths.ox.ac.uk}}

\begin{abstract}
We give a prescription for constructing a Lanczos potential for a
cosmological model which is a purely gravitational perturbation of a
Friedman-Lemaitre-Robertson-Walker (FLRW) space-time. For the
radiation equation of state, we find the Lanczos potential
explicitly via Fourier transforms. As an application, we follow up a
suggestion of Penrose \cite{Penrose} and propose a definition of
gravitational entropy for these cosmologies. With this definition,
the gravitational entropy initially is finite if and only if the
initial Weyl tensor is finite.
\end{abstract}
%Keywords: Lanczos potential; gravitational entropy; cosmology; perturbations.
\submitto{CQG}

\maketitle
%%%%%%%%%%%%%%%%%%%%%%%%%%%%%%%%%%%%%%%%%
\section{Introduction}
\label{sI}
%%%%%%%%%%%%%%%%%%%%%%%%%%%%%%%
While there is no generally agreed definition of gravitational
entropy in General Relativity (GR), it was conjectured some years
ago by Penrose \cite{Penrose}, that it should be related to the
clumping of matter, therefore to the degree of inhomogeneity and
anisotropy of a space-time and therefore associated with the Weyl or
conformal curvature. Specifically, Penrose suggested that a measure
for the gravitational entropy should involve an integral of a
quantity derived from the Weyl tensor, and that the definition of
the scalar product on the Hilbert space of one-particle states for
linearised GR in flat space might provide guidance \cite{Penrose}.
This definition, which we review below, is written in terms of
potentials for the (linearised) Weyl spinor.\footnote{For a
different approach to implementing Penrose's suggestion via graviton
number in FLRW cosmologies, see \cite{NO}. For more on potentials
  in linear theory and electromagnetism see e.g. chapter 6 of \cite{PenRin}.}

There have subsequently been several attempts to construct gravitational entropy measures
using polynomial invariants of the Weyl and Ricci tensors (see e.g.
\cite{Goode-Wainwright-iso,Bonnor85,Pelavas-Lake}) but
with no complete success. From the point of view of Penrose's suggestion these have the wrong `differential order' being constructed algebraically from the Weyl tensor rather than from potentials for it. Another approach has been to note that, via the Bianchi
identities, the Weyl tensor
is related to the density gradient,
which is a natural physical measure of inhomogeneity. There are
proposals for a definition of gravitational entropy based on density contrast
functions which are covariant, non-local, and work globally for dust
cosmologies, provided the initial singularity is isotropic \cite{Mena-Tavakol99.1,Hosoya}. This is encouraging, as part of the motivation of \cite{Penrose} was to connect the notion of low gravitational entropy to a restriction on the nature of the initial singularity.

It is unclear how any of these measures could relate to the established notion of black hole entropy or whether there should be a relation between the entropy in gravitational waves and black hole entropy at all.

Our purpose in this paper is to return to Penrose's suggestion and take it as far as possible, using the scalar product from linear theory in flat space to motivate a definition in curved space. This requires a potential for the Weyl tensor or Weyl spinor and we shall use the Lanczos potential
(\cite{Lanczos}, see also \cite{AE}, \cite{BC}). It is a general result of Illge \cite{Illge} that any spinor field with the symmetries of the Weyl spinor locally has a Lanczos potential which is (uniquely) determined by its value at a space-like hypersurface ${\mathcal{S}}$. Furthermore, for a vacuum space-time there exists a potential for the Lanczos potential, a second or super-potential for the Weyl spinor, again determined by its value at ${\mathcal{S}}$ \cite{Illge} (see also \cite{AE}).

Apart from Illge's result, there is no general prescription for obtaining a Lanczos potential for
a given spacetime. A
general expression for a Lanczos potential in the case of perfect
fluid spacetimes with zero shear and vorticity was given in \cite{NV}. More recently, this
result has been extended by Holgersson \cite{Holgersson}
to Bianchi I perfect-fluid spacetimes. There are
also several examples in the literature of Lanczos potentials for particular exact solutions, including G\"odel, Schwarzschild, Taub and Kerr
\cite{Berqvist,NV,Ovando-etal,Dolan,Edgar-etal}. Since there is freedom in the choice of the Lanczos potential, these examples are usually made subject to symmetry assumptions.

In this paper, we consider the Lanczos potential and superpotential
for linearly perturbed Friedman-Lemaitre-Robertson-Walker (FLRW)
spacetimes. We obtain wave equations for tensors defining invariant
parts of the Lanczos potential, and solve them by Fourier transforms
to give explicit solutions. Then, as an application, we closely
following Penrose's idea \cite{Penrose}, in order to propose a
measure of the gravitational entropy and apply it to linearly
perturbed FLRW cosmologies. The measure is defined as far as
possible in such a way as to carry over to more general situations.
Thus given a choice of `time', in the sense of a space-like
hypersurface, we construct a Lanczos potential with data at that
hypersurface, and a tensor which obeys the equations for a second
potential only at that hypersurface. Then we define a complex
structure at the hypersurface on the space of potentials. From this,
we construct a measure of gravitational entropy $S_g$ which is now a
functional of the (linearised) Weyl tensor.

Our definition is of course speculative. It has a reasonably good
motivation, but doesn't have any obvious monotonicity property. It
has at least one good property: the main positive result of this
part of the paper is that $S_g$ is finite at the initial singularity
only for those linearised Weyl tensors which are finite at the
initial singularity. Here the Weyl tensor initially is understood to
be finite if the metric (background plus perturbation) can be
rescaled so as to extend conformally through the singularity, in
other words if the initial singularity is still isotropic
\cite{Goode-Wainwright-iso}. If the initial Weyl tensor is singular,
in the sense understood here, then the initial gravitational entropy
is infinite. Thus finite initial gravitational entropy, as defined
here, requires finite initial Weyl tensor.

\medskip

The plan of the paper is as follows. To end this section we review
linearised GR, introducing the scalar product on the space of
solutions and explain how this motivates Penrose's suggestion. In
the next section, we review the $(1+3)$-formalism for cosmological
models and apply it to perturbations of FLRW cosmologies. Then in
Section 3, we obtain an expression for a Lanczos potential for these
perturbations, solving the wave equations which arise by Fourier
transforms. Finally in Section 4, we use the analysis from earlier
sections, together with a definition of complex structure on the
space of potentials, to suggest a definition of gravitational
entropy
 for these cosmologies.

\medskip

In linearised GR, we start by perturbing the flat metric $\eta_{ab}$ according to the equation
$$g_{ab}=\eta_{ab}+\Phi_{ab}.$$
For simplicity, we shall assume that the perturbation $\Phi_{ab}$ is
subject to the following gauge conditions:
\be
\Phi_a^{\;\;a}=0=\nabla^a\Phi_{ab}.
\label{lg4}
\ee
We obtain the linearised connection (in the sense of the perturbation of the Ricci
rotation coefficients) as the tensor $L_{abc}=L_{[ab]c}$ defined by
\be
L_{abc}=\nabla_{[a}\Phi_{b]c}.
\label{lg5}
\ee
Note that
\be
L_{ab}^{\;\;\;\;b}=0=\eta^{abcd}L_{abc}= \nabla^cL_{abc}.
\label{lg6}
\ee
With the convention
\[(\nabla_c\nabla_d-\nabla_d\nabla_c)V^a=R^a_{\;\;\;\;bcd}V^b,\]
the linearised Riemann tensor is
\be
R_{ab}^{\;\;\;\;cd}=-\nabla_{[a}L^{cd}_{\;\;\;\;b]}-\nabla^{[c}L_{ab}^{\;\;\;\;d]}.
\label{lg7}
\ee
The field equation is the linearisation of the Einstein vacuum equation and
is equivalent to the wave equation on $\Phi_{ab}$:
$$R_a^{\;\;c}:=R_{ab}^{\;\;\;\;cb}=\frac{1}{2}\Box \Phi_{ab}=0.$$
In spinors we write $\Phi_{ab}=\Phi_{ABA'B'}$ (not to be confused
with the Ricci spinor) and
\[L_{abc}=L_{ABCC'}\epsilon_{A'B'}+\;\;\mathrm{c.c.}\]
for a symmetric spinor $L_{ABCC'}$. Then (\ref{lg5}) and (\ref{lg7}), taking account of (\ref{lg4}), are
\bea
\nabla_C^{\;\;D'}\Phi_{ABC'D'}&=&-2L_{ABCC'},\label{lg8}\\
\nabla_D^{\;\;C'}L_{ABCC'}&=&\psi_{ABCD},\label{lg9} \eea
where $\psi_{ABCD}$ is the Weyl spinor.

We define the symplectic form on the space of solutions by
\be
\Omega(\Phi,\tilde{\Phi})=\int_{\mathcal{S}}(L_{abc}\widetilde{\Phi}^{bc}-\widetilde{L}_{abc}\Phi^{bc})dS^a,
\label{lg1}
\ee
where the integral is over a space-like hypersurface
${\mathcal{S}}$. Then this is independent of surface by virtue of
the field equation, and is also gauge-invariant. For the scalar
product on the Hilbert space of classical solutions, we need the
complex structure $J$ on the space of solutions, which is usually
defined by the splitting into positive and negative frequencies: if
$\Phi=\Phi_++\Phi_-$ is that splitting then
$$J\Phi=i(\Phi_+-\Phi_-),$$
and then the inner product is
\be
\langle \Phi, \Phi\rangle = \Omega(\Phi, J\Phi).
\label{lg2}
\ee
This can be seen to be positive-definite by writing it in terms of the Fourier transform.

This construction motivates Penrose's \cite{Penrose} suggested guide to a definition
of gravitational entropy: given a solution to a classical field theory, the coherent
state built on that classical state can be thought of as the most closely corresponding
quantum state; the expectation value of the number operator in the quantum state is
then a measure of the `number of particles' underlying the classical state, which in
turn is a measure of the entropy; but this expectation value is just the norm of the
classical state in the scalar product. If it were possible to find a definition like
(\ref{lg2}) above in a curved space-time, it would therefore be a candidate for a
definition of gravitational entropy. (Again, this might be just the entropy in gravitational waves.
 Black hole entropy could be another story.)

This is the idea we pursue here. We replace the linearised
connection $L_{abc}$ by the Lanczos potential, which by Illge
\cite{Illge} always exists satisfying (\ref{lg9}). The second or
superpotential $\Phi_{ab}$ derived from $L_{abc}$ always exists for
vacuum but we shall be concerned with cosmological solutions when
its existence is problematic, as is the correct definition of $J$.
On the other hand, we do not expect a definition which is
independent of surface or, equivalently, independent of time. Thus
our aim will be to mimic (\ref{lg2}) as close as possible but basing
the construction on a choice of hypersurface.

%%%%%%%%%%%%%%%%%%%%%%%%%%%%%%%%%%%%%%%%%%%%%%%%%%%%%%%%%%%%%%%%%%%%%%%%%%%%%%%%%%%%%%%%%%%%%%%
\section{1+3 Formalism and tensor perturbations of FLRW cosmologies}
\label{sF}
%%%%%%%%%%%%%%%%%%%%%%%%%%%%%%%%%%%%%%%%%%%%%%%%%%%%%%%%%%%%%%%%%%%%%%%%%%%%%%%%%%%%%%%%%%%%%%
We consider a space-time with a distinguished time-like direction given by the velocity
vector field $u^a$ of the fluid, and use the formalism of \cite{Ehlers,Ellis73,dynbook}, with $g_{ab}u^au^b=-1$. We introduce the tensor which, at each point, projects into the space orthogonal to
$u^a$ by
\begin{equation}
\label{ola1}
h_{ab}=g_{ab}+u_a u_b.
\end{equation}
Then
\begin{equation}
\label{ola2}
h_a^c h_c^b=h_a^b,~~~h_a^bu_b=0,~~~h_a^a=3.
\end{equation}
The covariant derivative of $u_a$ can be written, as usual, as
\begin{equation}
\label{ola3}
\label{dotu}
\nabla_a u_{b}=\frac{1}{3}{\theta}h_{ab}+\s_{ab}+\omega_{ab}-\dot{u}_au_b
\end{equation}
where
\begin{equation}
\label{trace}
\s_{ab}=\s_{(ab)};~~~\s^a_a=0;
~~~\s_{ab}u^b=0;~~~
\omega_{ab}=\omega_{[ab]};~~~\omega_{ab}u^b=0.
\end{equation}
%
%;~~~\s^2=\frac{1}{2}\s_{ab}\s^{ab}
Then $\dot u^a$ is the acceleration, $\omega_{ab}$ is the vorticity tensor,
$\s_{ab}$ the shear, and
$\theta$ the expansion.
The stress--energy tensor
for perfect fluids is
\begin{equation}
\label{ola8}
%(G_{ab}=)T_{ab}=\rho u_au_b+2q_{(a}u_{b)}+ph_{ab}+\pi_{ab}
T_{ab}=\rho u_au_b+ph_{ab}
\end{equation}
where $\rho$ is the energy density and $p$ the isotropic pressure of the
fluid.

We shall be principally concerned with the case of vanishing vorticity. Then the fluid flow is orthogonal to space-like hypersurfaces ${\cal{S}}_t$, which can be labelled by proper-time $t$ along the flow, $h_{ab}$ is the (Riemannian) metric on these hypersurfaces and its Levi-Civita covariant derivative, say $D_a$, is defined by projection: if $V_{b\ldots c}$ is a tensor orthogonal to $u^a$ on all indices then
\[D_aV_{b\ldots c}=h_a^dh_b^e\cdots h_c^f\nabla_dV_{e\ldots f}.\]
A useful operation below will be curl, defined for a symmetric tensor $X_{ab}$ orthogonal to $u^a$ by
\be
\label{curl}
(\curl~X)^{ab}:=\eta^{cd(a}D_c X^{b)}_{~d},
\ee
where $\eta_{abc}=\eta_{abcd}u^d$ is the volume form of ${\cal{S}}_t$, and $\eta_{abcd}$ is the space-time volume form. It will frequently be convenient to omit the brackets and write just $\curl~X^{ab}$ or $\curl~X_{ab}$.

The Weyl tensor can be decomposed into its electric and magnetic parts, $E_{ab}$ and $H_{ab}$ relative to
the velocity vector $u^a$ as

\begin{equation}
\label{ola23}
E_{ab}=C_{acbd}u^c u^d,~~~H_{ab}=C^*_{acbd}u^c u^d,
\end{equation}
where the dual $C^*_{acbd}$ is
\begin{equation}
\label{ola24}
C^*_{acbd}=\frac{1}{2}\eta_{ac}^{~~st} C_{stbd}.
\end{equation}
From their definition, $E_{ab}$ and $H_{ab}$ are symmetric, trace-free and orthogonal to $u^a$.

The Bianchi identities for the Weyl tensor in the case of a twist-free perfect fluid which also has vanishing acceleration can be written as the following system (see e.g. \cite{dynbook}):
\bea
\dot E^{ab}&=&-\theta E^{ab}-\frac{1}{2}(\rho+p)\sigma^{ab}+\curl
H^{ab}\nonumber\\
&&+3\sigma_c^{\;\;(a}E^{b)c}-\sigma_{cd}E^{cd}h^{ab},\label{bw1}\\
\dot H^{ab}&=&-\theta H^{ab}-\curl E^{ab} \nonumber\\
&&+3\sigma_c^{\;\;(a}H^{b)c}-\sigma_{cd}H^{cd}h^{ab},\label{bw2}\\
D_a E^{ab}&=& \eta^{bcd}\sigma_{ce}H^e_{\;\;d}+\frac{1}{3}D^b\rho,\label{bw3}\\
D_a H^{ab}&=& -\eta^{bcd}\sigma_{ce}E^e_{\;\;d},\label{bw4}
\eea
where the dot is $u^a\nabla_a$.

Now we use this formalism to consider perturbations of FLRW cosmologies which are purely gravitational. The background is conformally-flat, so that $E_{ab}=H_{ab}=0$ and the fluid-flow is geodesic, shear-free and twist-free so that
$$\dot u_a=\omega_{ab}=\sigma_{ab}=0.$$
We consider the FLRW metric $g_{ab}$ linearly perturbed with $\delta g_{ab}=\Phi_{ab}$. Following \cite{wein}, for purely gravitational perturbations we may consistently impose the  gauge conditions
\be
\label{g1}
\Phi_{ab}u^b=\Phi^a_{~a}=\nabla^a\Phi_{ab}=0.
\ee
For the linearised field equation, we characterise the perturbation as purely gravitational by requiring that the perturbation in the Ricci tensor in the form $R^b_a$ vanish:
\be
\label{e1}
\delta R^{b}_a=0.
\ee
This implies that $\delta\rho=\delta p=0$, and with the gauge conditions (\ref{g1}) also $\delta u^a=\delta u_a=0$, so that $\delta T^b_a=0$ for the stress-energy-momentum tensor.

For the perturbation in the kinematic quantities it easily follows,
for example in the formalism of \cite{dynbook}, that
$$\delta\theta=0=\delta\omega_{ab}=\delta\dot{u}_a$$
while, for the shear, we introduce the notation:
\be
\label{e2}
\Sigma_{ab}:=\delta\sigma_{ab}=\frac{1}{2}\dot\Phi_{ab}.
\ee
For the Weyl tensor, which is zero in the background, we find (from the equations in e.g. \cite{dynbook})
\begin{eqnarray}
\label{Epert}
E^{ab}&=&-\dot\Sigma^{ab}-\frac{2}{3}\theta~\Sigma^{ab},\\
\label{Hpert}
H^{ab}&=& \curl~\Sigma^{ab}.
\end{eqnarray}
Now the field equation (\ref{e1}) is
\be
\label{e3}
\Box\Phi_{ab}=\frac{2}{3}\rho\;\Phi_{ab},
\ee
%
% which can be written something like (check):
% %
% \[
% \ddot\Phi_{ab}+\frac{\dot a}{a} \dot\Phi_{ab}+(\nabla^2+k)\Phi_{ab}=0
% \]
%
(compare e.g. \cite{wein}). We note the following identities for trace-free, symmetric tensors $\chi_{ab}$ orthogonal to $u^a$:
\bea
D^a\curl~\chi_{ab}&=&0,\label{i1}\\
\nabla^a\chi_{ab}&=&D^a\chi_{ab},\label{i2}\\
(\nabla^a\chi_{ab})^\cdot&=&\nabla^a\dot\chi_{ab},\label{i3}\\
(\curl~\chi_{ab})^.&=&\curl~\dot \chi_{ab}-\frac{1}{3}\theta~\curl~\chi_{ab},\label{i4}\\
\curl~\curl~\chi_{ab}&=&-\ddot\chi_{ab}-\Box\chi_{ab}-\theta~\dot\chi_{ab}+(\rho-\frac{1}{9}\theta^2)\chi_{ab}.\label{i5}
\eea
Then, from (\ref{g1}), (\ref{e2}), (\ref{i2}) and (\ref{i3})
\be
\label{i6}
D^a\Sigma_{ab}=0,
\ee
and from (\ref{e2}) and (\ref{e3}) we calculate
\be
\Box \Sigma_{ab}=\frac{2}{3}\theta~\dot{\Sigma}_{ab}+(\frac{1}{6}\rho-\frac{3}{2}p+\frac{1}{3}\theta^2)\Sigma_{ab}.
\label{i7}
\ee

It is now easy to check that, neglecting second-order terms, (\ref{bw1})-(\ref{bw4}) are satisfied with $E_{ab}$ and $H_{ab}$ as in (\ref{Epert}) and (\ref{Hpert}). Specifically (\ref{bw2}) follows from (\ref{i4}), (\ref{bw3}) from (\ref{i3}) and (\ref{i6}), (\ref{bw4}) from (\ref{i1}), and finally (\ref{bw1}), which is the hardest, from (\ref{e3}), (\ref{i5}), (\ref{i7}) and the Raychaudhuri equation:
\be
\dot{\theta}+\frac{\theta^2}{3}+\frac{1}{2}(\rho+3p)=0.
\label{Ray}\ee

%%%%%%%%%%%%%%%%%%%%%%%%%%%%%%%%%%%%%%%%%%%%%%%%%%%%%%%%%%%%%%%%%%%%%%%%%%%%%%%%%%%%%%%%%%%%%%%%
\section{The Lanczos potential}
\label{sLP}
%%%%%%%%%%%%%%%%%%%%%%%%%%%%%%%%%%%%%%%%%%%%%%%%%%%%%%%%%%%%%%%%%%%%%%%%%%%%%%%%%%%%%%%%%%%%%%%%
The Lanczos potential is a tensor $L_{abc}=-L_{bac}$, connected to the Weyl tensor by the equation:
\be
\label{l1}
C_{ab}^{\;\;\;\;cd}=
-\nabla^{[c}L_{ab}^{\;\;\;\;d]}-\nabla_{[a}L^{cd}_{\;\;\;\;b]}-\mathrm{traces},
\ee
which should be compared with (\ref{lg7}) (in general, we follow \cite{Holgersson} but our definition of $L_{abc}$ is twice the usual definition, in order to maintain (\ref{lg5})).
There is gauge freedom in $L_{abc}$ satisfying (\ref{l1}), which can be reduced by imposing the Lanczos gauge conditions:
\[
L_{ab}^{\;\;\;\;b}=0=\eta^{abcd}L_{abc}=\nabla_cL_{ab}^{\;\;\;\;c},
\]
the same conditions as in (\ref{lg6}). When these are imposed, the `-traces' term in (\ref{l1}) is $-2\delta_{[a}^{[\;\;c}Q_{b]}^{\;\;d]}$ where
\[Q_{ac}=\nabla^bL_{abc},\]
which is symmetric and trace-free by virtue of the gauge-conditions
on $L_{abc}$. This term vanishes in the Minkowski space version of
the theory described in Section~1, but doesn't necessarily vanish in
curved space.

The algebraic gauge conditions ensure that $L_{abc}$ can be expressed in terms of a symmetric spinor field $L_{ABCC'}$ as
\[L_{abc}=L_{ABCC'}\epsilon_{A'B'}+\bar{L}_{A'B'C'C}\epsilon_{AB},\]
and the differential gauge condition then implies
\be
\label{l2}
\nabla^{CC'}L_{ABCC'}=0.
\ee
Now (\ref{l1}) takes the spinor form
\be
\label{l3}
\nabla_D^{\;\;C'}L_{ABCC'}=\psi_{ABCD},
\ee
just as in (\ref{lg9}) but where $\psi_{ABCD}$ is now the full (nonlinear) Weyl spinor. There is no need to symmetrise in (\ref{l3}) because of (\ref{l2}). Illge \cite{Illge} shows that (\ref{l3}) has a unique solution given $L_{ABCC'}$ on a space-like surface, but one cannot in general find the second potential as in (\ref{lg8}) as this equation has a curvature obstruction from the Ricci tensor: given (\ref{lg9}), (\ref{lg8}) implies
\be
\varphi^{C'D'E}_{\;\;\;\;\;\;\;\;\;\;(A}\Phi_{B)EC'D'}=0,
\label{obs}
\ee
where $\varphi_{ABA'B'}$ is the Ricci spinor.

Holgersson \cite{Holgersson} gave a useful decomposition of the
Lanczos potential into irreducible parts in the (1+3)-formalism as follows:
\begin{equation}
\label{split} L_{abc}=2u_{[a}A_{b]}u_c-A_{[a}h_{b]c}-
2u_{[a}C_{b]c}+\eta_{ab}^{~~d}S_{dc}+u_{[a}\eta_{b]cd}P^d-u_c\eta_{abd}P^d,
\end{equation}
where $A_a$ and $P_a$ are orthogonal to $u^a$ and $S_{ab}$ and
$C_{ab}$ are trace-free, symmetric and orthogonal to $u^a$. This
gives sixteen components for $L_{abc}$ (three each for $A$ and $P$;
five each for $S$ and $C$) which agrees with the eight complex
components for $L_{ABCC'}$. Holgersson \cite{Holgersson} also gave
formulae for the electric and magnetic parts of the Weyl tensor
which are useful below.

We want to calculate a Lanczos potential for a perturbed FLRW spacetime with some given perturbation $\Phi_{ab}$ in the metric, as considered in Section~\ref{sF}. Since the perturbation is characterised by a trace-free, symmetric tensor orthogonal to $u^a$, we seek a Lanczos potential as in (\ref{split}) with vector parts zero: $A_a=P_a=0$. Then from (\ref{split}) and (\ref{l1}), or quoting from \cite{Holgersson} we find
\bea
\label{weyl1}
E_{ab}&=&\frac{1}{2}(\curl~S_{ab}-\dot
C_{ab}),\\
\label{weyl2}
H_{ab}&=&\frac{1}{2}(\curl~C_{ab}+\dot
S_{ab}).
\eea
Equating these to (\ref{Epert}) and (\ref{Hpert}) we have equations for $C_{ab}$ and
$S_{ab}$.
Now, if a superpotential $\phi_{ab}$ also existed for all times with
\be
L_{abc}=\nabla_{[a}\phi_{b]c}
\label{f7}
\ee
then, from (\ref{split}), we would have
\bea
C_{ab}&=& \frac{1}{2}(\dot \phi_{ab}+\frac{\theta}{3}\phi_{ab}),\label{CS1}\\
S_{ab}&=& \frac{1}{2}\curl~\phi_{ab},\nonumber
\eea
but this is incompatible with the Bianchi identities (\ref{bw1})-(\ref{bw4}).
This incompatibility is a consequence of the obstruction (\ref{obs}).
 However, (\ref{CS1}) suggests another ansatz, namely
\bea
\label{CS}
C_{ab}&=&\frac{1}{2}(\psi_{ab}+\frac{\theta}{3} \phi_{ab})\nonumber\\
S_{ab}&=&\frac{1}{2}\curl~\phi_{ab}
\eea
in terms of another unknown tensor $\psi_{ab}$. (This is simply an
ansatz, in that we express two unknown tensors, $C$ and $S$, in
terms of two other unknown tensors,
 $\phi$ and $\psi$; the justification of the ansatz is the
  simplification which results, for example in (\ref{f8}) and (\ref{f9})
  below).

Using (\ref{i4}) we find from (\ref{weyl2}) and (\ref{CS})
\bea
H_{ab}=\frac{1}{4}\curl~(\dot\phi+\psi)_{ab}.
\eea
Comparing this equation with (\ref{Hpert}) we can choose
\begin{equation}
\label{constraint}
\Sigma_{ab}=\frac{1}{4}(\dot \phi_{ab}+\psi_{ab}),
\end{equation}
so that $\psi_{ab}$ is known once $\phi_{ab}$ has been found. Then, using (\ref{weyl1})
\begin{equation}
\label{electric-l}
E_{ab}=\frac{1}{4}(-\dot\psi_{ab}-\frac{\dot\theta}{3}\phi_{ab}-\frac{\theta}{3}\dot\phi_{ab}+\curl~\curl~\phi_{ab}),
\end{equation}
and combining this with (\ref{Epert}), (\ref{i5}) and (\ref{constraint}) we get
\begin{equation}
\label{constraint2}
\Box \phi_{ab}+\frac{4}{3}\theta\dot\phi_{ab}+(\frac{\dot
\theta}{3}+\frac{\theta^2}{9}-\rho)\phi_{ab}=\frac{8}{3}\theta\Sigma_{ab},
\end{equation}
which is a wave equation for $\phi_{ab}$. Note that this
wave-equation is \emph{not} (\ref{e3}). In fact, if we introduce
$X_{ab}=\phi_{ab}-\Phi_{ab}$ then $X_{ab}$ satisfies
\be
\Box X_{ab}+\frac{4}{3}\theta\dot{X}_{ab}+(\frac{\dot
\theta}{3}+\frac{\theta^2}{9}-\rho)X_{ab}=\frac{1}{2}(\rho+p)\Phi_{ab},
\label{f6}
\ee
%%%
%%%
using the Raychaudhuri equation (\ref{Ray}) again. Equation (\ref{f6}) is not satisfied by zero, so that $\phi_{ab}$ cannot be taken to be the perturbed metric $\Phi_{ab}$. For later use, we note that, in terms of $X_{ab}$ and $\Phi_{ab}$:
\bea
\phi_{ab}&=&\Phi_{ab}+X_{ab},\label{f8}\\
\psi_{ab}&=&\dot{\Phi}_{ab}-\dot{X}_{ab}.\label{f9}
\eea
Given initial
data $(\phi_{ab}({\bf x},t_0),\dot\phi_{ab}({\bf x},t_0))$ or equivalently $(\phi_{ab}({\bf x},t_0),\psi_{ab}({\bf x},t_0))$, in terms of spatial coordinates ${\bf x}$ at some time $t_0$, a solution to (\ref{constraint2}) exists and is unique. We therefore have a
complete prescription to determine a unique $L_{abc}$ for linearly perturbed FLRW, subject to this data. We can achieve (\ref{CS1}), and therefore (\ref{f7}), at a given instant by choosing the data to be
\bea
\phi_{ab}({\bf x},t_0)&=&\Phi_{ab}({\bf x},t_0)\label{dat}\\
\dot{\phi}_{ab}({\bf x},t_0)&=&\dot{\Phi}_{ab}({\bf x},t_0)\nonumber
\eea
at that instant, or equivalently
\be
X_{ab}({\bf x},t_0)=0=\dot{X}_{ab}({\bf x},t_0),
\label{dat2}
\ee
 but this won't then be true at other times.

We summarize our results so far in
the following proposition:
\\\\
{\bf Proposition:}  {\em Given  a perturbed FLRW
spacetime and a choice of time $t_0$, a Lanczos potential $L_{abc}$, in the Lanczos gauge, may be uniquely specified by
(\ref{split}) with (\ref{CS}), (\ref{constraint}) and (\ref{constraint2}), subject to the data (\ref{dat}).
 We may define a superpotential $\phi_{ab}$ such that (\ref{f7}) holds at $t_0$ but this will not hold at other times.}
\\\\
To obtain explicit solutions for $\phi$ and $X$, we first recall
some details of the FLRW metrics.
 For simplicity, we shall assume $k=0$, the spatially-flat case, so that the metric is
\[g=R(t)^2(dx^2+dy^2+dz^2)-dt^2,\]
or, introducing conformal time $\tau$,
\[g=\tilde{R}(\tau)^2(dx^2+dy^2+dz^2-d\tau^2),\]
where $R(t)=\tilde{R}(\tau)$ and $dt/R(t)=d\tau$. The overdot will consistently stand for $d/dt$, and $d/d\tau$ will always be written explicitly.

We choose, as spatial coordinates, ${\bf x}=(x^{\bf i})$ for ${\bf i}=1,2,3$, and note that $dS=R^3d^3{\bf x}$ is the volume element on the hypersurfaces of constant $t$. We shall assume an equation of state of the form $p=(\gamma-1)\rho$, and then $\theta=3\dot R/R$, while the conservation equation implies that $\rho=\rho_0 R^{-3\gamma}$ for constant $\rho_0$. The Friedmann equation reduces to
\be
\dot{R}^2=\frac{1}{3}\rho R^2,\label{fr1}\ee
(with the convention that $8\pi G/c^2=1$) and, without loss of generality, the solution is $R=t^n$ where $n=\frac{2}{3\gamma}$.

In coordinate components, for a tensor $\chi_{ab}$ orthogonal to $u^a$ the d'Alembertian is
\be
\Box \chi_{\bf{ij}}=\frac{1}{R^2}\Delta_0 \chi_{\bf{ij}}-\ddot{ \chi}_{\bf{ij}}+\frac{\dot R}{R}\dot{ \chi}_{\bf{ij}}+2(\frac{\ddot{R}}{R}+\frac{\dot{R}^2}{R^2}) \chi_{\bf{ij}}.\label{DA}\ee
where $\Delta_0$ is the flat Laplacian in ${\bf x}$, and for the time-evolution we find, in components:
\[(u^c\nabla_c \chi)_{\bf{ij}}=\dot{ \chi}_{\bf{ij}}-\frac{2\dot{R}}{R} \chi_{\bf{ij}}.\]
Define the Fourier transform $\hat{\Phi}_{{\bf ij}}(t, {\bf q})$ of the coordinate components $\Phi_{{\bf ij}}(t, {\bf x})$ of $\Phi_{ab}$ in the usual way as
\be
\hat{\Phi}_{{\bf ij}}(t, {\bf q})=\frac{1}{(2\pi)^{3/2}}\int_{{\bf R}^3}\Phi_{{\bf ij}}(t, {\bf x})\exp(i{\bf{q\cdot x}})d^3{\bf x},
\label{F1}
\ee
then, suppressing indices for clarity, (\ref{e3}) with the aid of (\ref{DA}), becomes
\[\ddot{\hat\Phi}-\frac{\dot R}{R}\dot{\hat\Phi}-(\frac{2\ddot{R}}{R}-\frac{|{\bf q}|^2}{R^2})\hat\Phi=0,\]
which, as a check, is equation 15.10.39 of \cite{wein} (when making the comparison, recall that
\[\frac{\ddot{R}}{R}=-\frac{1}{6}(\rho+3p)\]
with our conventions). Following \cite{wein}, substitute
\be
\hat\Phi_{{\bf ij}}=\tau^\alpha(h^+_{{\bf ij}}({\bf q}) \hat{H}^+(\tau,{\bf q})+h^\times_{{\bf ij}}({\bf q}) \hat{H}^\times(\tau,{\bf q})),
\label{P1}
\ee
where $\alpha=\frac{1+n}{2(1-n)}$ and $h^+_{{\bf ij}}({\bf q})$ and $h^\times_{{\bf ij}}({\bf q})$ represent the two polarisation states i.e. they are symmetric, trace-free matrices of a standard form, orthogonal to ${\bf q}$ and suitably normalised. Then (\ref{P1}) gives
\[\frac{d^2\hat H}{d\tau^2}+\frac{1}{\tau}\frac{d\hat H}{d\tau}+(|{\bf q}|^2-\frac{\nu^2}{\tau^2})\hat H=0,\]
where $\nu=\frac{(3n-1)}{2(1-n)}$, which is Bessel's equation of order $\nu$ in $|{\bf q}|\tau$, so that each of $\hat H^+$ and $\hat H^\times$ is a ${\bf q}$-dependent linear combination of Bessel functions $J_{\pm\nu}(|{\bf q}|\tau)$.

Now we need the Fourier transform of (\ref{f6}), in coordinate components, taking account of (\ref{fr1}), and again suppressing indices for clarity. This is
\be
\ddot{\hat X}-\frac{5\dot{R}}{R}\dot{\hat X}-(\frac{3\ddot{R}}{R}-\frac{9\dot{R}^2}{R^2}-\frac{|{\bf q}|^2}{R^2})\hat X=-\frac{1}{2}(\rho+p)\hat\Phi.
\label{X1}
\ee
Analogously to (\ref{P1}), put
\be
\hat X_{{\bf ij}}=\tau^\beta(h^+_{{\bf ij}}({\bf q}) \hat{G}^+(\tau,{\bf q})+h^\times_{{\bf ij}}({\bf q}) \hat{G}^\times(\tau,{\bf q})),
\label{X11}
\ee
with $\beta=\frac{5n+1}{2(1-n)}$ to find
\be
\frac{d^2\hat G^+}{d\tau^2}+\frac{1}{\tau}\frac{d\hat G^+}{d\tau}+(|{\bf q}|^2-\frac{1}{4\tau^2})\hat G^+=\frac{-n}{(1-n)^2}\tau^{\alpha-\beta-2}\hat H^+,
\label{X2}
\ee
%
%Note $\alpha-\beta-2 =\frac{-2}{1-n}$.
and similarly for $G^\times$ in terms of $H^\times$.

We may solve (\ref{X2}) by variation of parameters: suppose $t=t_0$ corresponds to $\tau=a$, then we want $\hat G(a)=\dot{\hat G}(a)=0$ by (\ref{dat2}). The homogeneous equation is Bessel's equation with $\nu=1/2$, and the solution for $\hat G^+$ is
\be
\hat G^+(\tau)=\frac{-n\tau^{-1/2}}{(1-n)^2|{\bf q}|}\int_a^\tau \sigma^{\alpha-\beta-3/2}\sin(|{\bf q}|(\tau-\sigma))\hat H^+(\sigma)d\sigma,
\label{X3}
\ee
and similarly for $\hat G^\times$.

\medskip

To summarise: (\ref{P1}), (\ref{X11}) and (\ref{X3}) determine $X$ and $\Phi$ and then (\ref{f8}) and (\ref{f9}) determine $\phi$ and $\psi$; finally (\ref{CS}) and (\ref{split}) determine the Lanczos potential.

%%%%%%%%%%%%%%%%%%%%%%%%%%%%%%%%%%%%%%%%%%%%%%%%%%%%%%%%%%%%%%%%%%%%%%%%%%%%%%%%%%%%%%%%%%%%%%%%%%%
\section{Gravitational Entropy for perturbed FLRW}
%%%%%%%%%%%%%%%%%%%%%%%%%%%%%%%%%%%%%%%%%%%%%%%%%%%%%%%%%%%%%%%%%%%%%%%%%%%%%%%%%%%%%%%%%%%%%%%%%%%%%%
%%%%%%%%%%%%%%%%%%%%%%%%%%%%%%%%%%%%%%%%%%%%%%%%%%%%%%%%%%%%%%%%%%%%%%%%%%%%%%%%%
We want to recapitulate as much as possible of the argument in Section 1, using the Lanczos potential in place of the $L$ used there. In place of (\ref{lg1}), given two solutions of (\ref{e3}), the Einstein equations for gravitational perturbations of FLRW, we shall define
\be
\Omega(\Phi_{(1)},\Phi_{(2)})=\int_{{\mathcal{S}}_{t}}
\left(L^{(1)}_{abc}\phi^{(2)bc}-L^{(2)}_{abc}\phi^{(1)bc}\right )
u^a dS,
\label{OO}
\ee
with $\phi^{(i)}_{ab}$ related to $L^{(i)}_{abc}$ as in (\ref{f7}) at a particular choice of $t_0$. Then from (\ref{split}) and (\ref{CS}) we find
\be
\Omega(\Phi_{(1)},\Phi_{(2)})=\frac{1}{2}\int_{{\mathcal{S}}_{t}}
\left(\psi^{(1)}_{ab}\phi^{(2)ab}-\phi^{(1)}_{ab}\psi^{(2)ab}\right )dS.
\label{O1}
\ee
Unlike the case of linear
theory, this integral is not independent of time, since
\[\nabla^a
(L^{(1)}_{abc}\phi^{(2)bc}-L^{(2)}_{abc}\phi^{(1)bc})\ne 0.
\]
 From (\ref{O1}), using (\ref{f8}) and (\ref{f9})
\bea
\Omega(\Phi_{(1)},\Phi_{(2)})&=&\frac{1}{2}\int_{{\mathcal{S}}_{t}}
(\dot{\Phi}_{(1)ab}\Phi_{(2)}^{ab} -\dot{\Phi}_{(2)ab}\Phi_{(1)}^{ab})~dS\nonumber\\
&&-\frac{1}{2}\int_{{\mathcal{S}}_{t}}( X_{(1)ab}\Phi_{(2)}^{ab}- X_{(2)ab}\Phi_{(1)}^{ab})^\cdot dS\nonumber\\
&&-\frac{1}{2}\int_{{\mathcal{S}}_{t}}(\dot{ X}_{(1)ab} X_{(2)}^{ab}-\dot{ X}_{(2)ab} X_{(1)}^{ab})~dS.
\label{f10}
\eea
Define the current
\[j_\Phi^a:=\Phi^{bc}_{(1)}\nabla^a\Phi_{(2)bc}-\Phi^{bc}_{(2)}\nabla^a\Phi_{(1)bc},\]
then by (\ref{e3}) $j_\Phi^a$ is conserved:
\[\nabla_aj_\Phi^a=0.\]
Therefore the integral
\be
\int_{{\mathcal{S}}_t} u_a j_\Phi^{\;a}dS=-\int_{{\mathcal{S}}_t}(\dot{\Phi}_{(1)ab}\Phi_{(2)}^{ab} -\dot{\Phi}_{(2)ab}\Phi_{(1)}^{ab})~dS
\label{sym}
\ee
is constant in time i.e. the first integral in (\ref{f10}) is necessarily constant in time, but the others won't be. (It might be argued that one should use just (\ref{sym}) as the definition of the symplectic form, precisely because it {\emph{is}} independent of time; however this is a definition which would not be available in more general settings, whereas (\ref{OO}) would.)

Our strategy now is the following: we seek a definition of gravitational entropy based on (\ref{lg2}); we don't have a symplectic form which is independent of time, nor do we expect to obtain a definition of entropy which is independent of time; however we do have a unique definition of first and second potential for the Weyl tensor given a choice of time; therefore we shall make a natural choice of complex structure, given a choice of time, and use (\ref{lg2}) at that time to define the entropy.  By using (\ref{O1}) precisely at $t_0$ we simplify (\ref{f10}) greatly, since $X_{ab}$ vanishes there.

Let us for simplicity restrict to the radiation equation of state, so that $\gamma=4/3$, and then $n=1/2$, $\alpha=3/2$, $\beta=7/2$ and $\nu=1/2$, and we can take $R=t^{1/2}=\tau/2$ and $\rho R^4=3/4$. The Bessel functions $J_{\pm 1/2}$ can be written in terms of elementary functions, and as solutions for $\hat{H}^{(i)}$ we can take
\be
\hat H^{(i)}=\tau^{-1/2}(A^{(i)}({\bf q})\sin(|{\bf q}|\tau)+B^{(i)}({\bf q})\cos(|{\bf q}|\tau)).\label{Q6}
\ee
Reality of $\Phi_{ab}$ implies that
\be
\bar{\hat{\Phi}}(t, {\bf q})=\hat{\Phi}(t,-{\bf q})
\label{F2}
\ee
and then from (\ref{P1}) and (\ref{Q6}) the same holds for $A^{(i)}$ and $B^{(i)}$.

Note that $A$ parametrises a `growing mode' and $B$ a `decaying mode' in the standard terminology, and note also that a growing mode is a perturbation of the conformal metric which is finite at the initial singularity while a decaying mode is not (essentially because for the former $\Phi=O(t)$ as $t\rightarrow 0$, just like the unperturbed spatial metric, while for the latter $\Phi=O(t^{1/2})$, which diverges by comparison with the spatial metric). Thus perturbations with nonzero $B$ have infinite initial Weyl tensor while perturbations with $B$ zero but $A$ nonzero have finite initial Weyl tensor. This will be important below.

Introduce
\be
K({\bf q})=A^{(1)}({\bf q})B^{(2)}(-{\bf q})-B^{(1)}({\bf q})A^{(2)}(-{\bf q})
\label{D1}
\ee
then the first integral in (\ref{f10}) is
\be
\frac{1}{2}\int_{{\mathcal{S}}_{t}}
(\dot{\Phi}_{(1)ab}\Phi_{(2)}^{ab} -\dot{\Phi}_{(2)ab}\Phi_{(1)}^{ab})~dS
=\frac{2}{(2\pi)^{3/2}}\int |{\bf q}|K({\bf q})d^3{\bf q},
\label{f12}
\ee
which, as expected, is constant in time.

For the complex structure on the space of solutions of (\ref{e3}), we follow the discussion in \cite{AM}. There, for Klein-Gordon fields, the symplectic form is taken to be
\[\Omega((\phi, \pi), (\tilde{\phi}, \tilde{\pi}))=\int_{\mathcal{S}_t}(\pi\tilde{\phi}-\phi\tilde{\pi})dS,\]
with $\pi=\dot{\phi}$, and the complex structure is defined by
\[J\phi=(\Theta)^{-1/2}\pi\qquad ;\qquad J\pi=-(\Theta)^{1/2}\phi.\]
Here $\Theta=-D_aD^a$, where $D_a$ is the intrinsic covariant
derivative on ${\mathcal{S}}_t$. We seek to follow this
prescription, with (\ref{O1}) as the symplectic form so that $\psi$
then takes the role of $\pi$. Also $\Theta=-R^{-2}\Delta_0$ where,
as before, $\Delta_0$ is the flat, three-dimensional Laplacian. In
terms of the Fourier transform, $\Theta=|{\bf q}|^2R^{-2}$ so that
$J$ becomes
\be
J\hat\phi=\frac{R}{|{\bf q}|}\hat\psi\qquad ; \qquad J\hat{\psi}=-\frac{|{\bf q}|}{R}\hat\phi.\label{J2}\ee
We use (\ref{P1}) and (\ref{Q6}) to translate (\ref{J2}) into an
action on $A$ and $B$, all evaluated at $\tau=a$. First for the
relation of $\hat\phi$ and $\hat\psi$ to $A$ and $B$ from (\ref{P1})
and (\ref{Q6}) we have
\begin{displaymath}\left(\begin{array}{c}\hat\phi\\\hat\psi\end{array}\right)=\left(\begin{array}{cc}M_{11}&
M_{12}\\
M_{21}& M_{22}\end{array}\right)\left(\begin{array}{c}A\\B\end{array}\right)
\end{displaymath}
where
\bean
M_{11}&=&a\sin a|{\bf q}|\\
M_{12}&=&a\cos a|{\bf q}|\\
M_{21}&=&\frac{2}{a}(\sin a|{\bf q}|+a|{\bf q}|\cos a|{\bf q}|)\\
M_{22}&=&\frac{2}{a}(\cos a|{\bf q}|-a|{\bf q}|\sin a|{\bf q}|)
\eean
Then, conjugating $J$ defined by (\ref{J2}) with $M$ and introducing
$\alpha:=a|{\bf q}|$, we find the action of $J$ on $A$ and $B$ to be
\begin{displaymath}\left(\begin{array}{c}JA\\JB\end{array}\right)=\left(\begin{array}{cc}J_{11}&
J_{12}\\
J_{21}& J_{22}\end{array}\right)\left(\begin{array}{c}A\\B\end{array}\right)
\end{displaymath}
where
\bea
J_{11}&=&-\frac{1}{\alpha}\cos 2\alpha-\frac{1}{2\alpha^2}\sin 2\alpha, \label{J3}\\
J_{12}&=&-\frac{1}{\alpha^2}(\cos\alpha-\alpha\sin\alpha)^2-\cos^2\alpha,\nonumber\\
J_{21}&=&\frac{1}{\alpha^2}(\sin\alpha+\alpha\cos\alpha)^2+\sin^2\alpha,\nonumber\\
J_{22}&=&-J_{11}.\nonumber
\eea
We note the behaviour as
 $a\rightarrow 0$, when
\bea
J_{11}&=&-\frac{2}{\alpha}+O(\alpha),\label{J4}\\
J_{12}&=&-\frac{1}{\alpha^2}+O(1),\nonumber\\
J_{21}&=&4+O(\alpha^2),\nonumber\\
J_{22}&=&-J_{11}.\nonumber
\eea
In (\ref{D1}), following the analogy of (\ref{lg2}), we set $A^{(2)}=JA^{(1)}$ and $B^{(2)}=JB^{(1)}$, to find
\bea
K({\bf q}, a)&=&J_{21}A^{(1)}({\bf q})\overline{A^{(1)}({\bf q})}+J_{22}A^{(1)}({\bf q})\overline{B^{(1)}({\bf q})}\nonumber\\
&&-J_{11}B^{(1)}({\bf q})\overline{A^{(1)}({\bf q})}-J_{12}B^{(1)}({\bf q})\overline{B^{(1)}({\bf q})}.\label{J6}
\eea
where the choice of $J$ has introduced an explicit dependence on $a$. Substituting this into (\ref{f12}) gives our definition of gravitational entropy at proper time $t_0$, corresponding to conformal time $\tau=a$:
\be
S_g(\Phi;t_0)=\frac{2}{(2\pi)^{3/2}}\int |{\bf q}|K({\bf q},a)d^3{\bf q},
\label{J5}
\ee
with $K({\bf q},a)$ as in (\ref{J6}).

\medskip

We deduce the following properties of our definition:

\begin{itemize}
\item The entropy at time $t_0$ is determined by the Weyl tensor at $t_0$, but not locally (since e.g. the operator $(\Theta)^{-1/2}$ in (\ref{J2}) is non-local). In particular, it is not the integral of a scalar invariant of curvature.
\item From the specific form (\ref{J3}) of $J$ (though it follows more generally) $S_g$ can be seen to be positive definite.

\item However, for small $a$ we note from (\ref{J4}) that if $B\neq 0$ then $K=O(a^{-2})$,
which diverges on the approach to the initial singularity, while if
$B=0$ but $A\neq 0$ then $K=O(1)$, which is finite. In other words,
with this definition the initial gravitational entropy is finite if
the initial Weyl tensor is finite, and infinite otherwise. This is
an important success for the definition.

\item Necessarily, for a decaying mode the gravitational entropy will decay from its infinite initial value and tend to a positive constant in the remote future (for this, from (\ref{J6}), we note that $J_{12}$ in (\ref{J3}) tends to a constant as $t\rightarrow\infty$). But $J_{12}$ {\emph{isn't}} monotonic in time, which makes it unlikely that $S_g$ is.

\item From (\ref{J3}),  $J_{21}$ runs from a value of 4 at $\alpha=0$ to a limit of 1 at large $\alpha$ and is also not monotonic (though it is positive). Thus for a purely growing mode, the gravitational entropy runs from a finite initial value to a finite but lower final value, and again there is no reason to expect it to be monotonic. This is perhaps less successful for the definition, but this is after all a result from linear theory, though in a non-flat background - without nonlinearity, there is no gravitational clumping.
\end{itemize}
The definition has been framed in such a way that it should extend to more general situations, and it remains to be seen whether the property of being initially finite only for an isotropic singularity persists.
\section*{Acknowledgements}%\nonumber
PT acknowledges useful discussions with Roger Penrose and Abhay
Ashtekar, and FM with Brian Edgar, Roger Penrose and Reza Tavakol as
well as support from CMAT - Univ. Minho and from FCT (Portugal)
through grant SFRH/BPD/12137/2003.
\\\\

%%%%%%%%%%%%%%%%%%%%%%%%%%%%%%%%%%%%%%%%%%%%%%%%%%%%%%%%%%%%%%%%%%%%%%%%%%%%%%%%%%%%%%%%%%%%%%%%%%%%%%%%%%%%%%

%_____________________________________________________________________________
\end{document}